\documentclass[twocolumn,english]{revtex4}
\usepackage[T1]{fontenc}
\usepackage[latin1]{inputenc}
\usepackage{graphicx}
\pdfoutput=1

\makeatletter

\usepackage{babel}
\makeatother
\begin{document}

\title{Pinwheel stability, pattern selection and the geometry of visual
space }

\begin{abstract}
It has been proposed that the dynamical stability of topological defects
in the visual cortex reflects the Euclidean symmetry of the visual
world. We analyze defect stability and pattern selection in a generalized
Swift-Hohenberg model of visual cortical development symmetric under
the Euclidean group E(2). Euclidean symmetry strongly influences the
geometry and multistability of model solutions but does not directly
impact on defect stability. 
\end{abstract}

\author{Michael Schnabel$^{1,2}$, Matthias Kaschube$^{1,2,3,4}$ and Fred
Wolf$^{1,2}$}

\affiliation{$^{1}$Max Planck Institute for Dynamics and Self-Organization, Goettingen,
Germany, $^{2}$Bernstein Center for Computational Neuroscience, University
of Goettingen, Goettingen, Germany, $^{3}$Lewis-Sigler Institute,
$^{4}$Physics Department, Princeton University, Princeton, New Jersey,
USA}

\maketitle
When an object in the visual world is rotated about the axis of sight
the orientations of distant contours change in a coordinated manner,
leaving relative orientations of distant contours invariant. In the
visual cortex of the brain, it has been predicted that this geometrical
property of visual space imposes the so called shift-twist symmetry
on joint representations of contour position and contour orientation
\cite{Bressloff:2001}. This symmetry requires the equivariance of
dynamical models for these representations under the Euclidean group
E(2). In particular, it has been hypothesized that shift-twist symmetry
stabilizes topological pinwheel-defects in models for the emergence
of orientation selectivity during cortical development \cite{Mayer:2002,Lee:2003,Bressloff:2005}.
Pinwheel defects are singular points in the visual cortex around which
each orientation is represented exactly once \cite{Swindale:1987b,Bonhoeffer:1991,Ohki:2006}.
They are initially generated in the visual cortex at the time of eye
opening \cite{White:2001b,Chapman:1996}. This process has been theoretically
explained by spontaneous symmetry breaking \cite{Swindale:1982b,Wolf:1998}.
Why in the brain, pinwheels remain present at all developmental stages,
although they are dynamically unstable in many models of visual cortical
development \cite{Wolf:1998,Koulakov:2001,Mayer:2002,Cho:2004,Lee:2003,Wolf:2005b}
and in analogous physical systems \cite{Cross:1993,Vilenkin:1994,Bodenschatz:2000},
remains unclear. 

Previous studies \cite{Wolf:1998,Koulakov:2001,Mayer:2002,Lee:2003,Cho:2004,Wolf:2005b}
found pinwheels dynamically unstable only in models exhibiting an
E(2)xU(1) symmetry, which is higher than the E(2) symmetry of visual
perceptual space. In contradistinction, models exhibiting only Euclidean
symmetry have been shown to exhibit stable pinwheels \cite{Mayer:2002,Lee:2003,Thomas:2004}
suggesting that the stabilization of pinwheel defects may be closely
related to the 'reduced' Euclidean symmetry. There are, however, also
other scenarios that predict the emergence of stable pinwheels with
higher than Euclidean symmetry \cite{Wolf:2005b,Koulakov:2001} leaving
it hard to judge to actual role of Euclidean symmetry in orientation
map development. 

Here we analyze the impact of Euclidean symmetry on pattern selection,
i.e. the question of whether stable pinwheel arrangement exist and
what their geometric organization is. We construct a generalized Swift-Hohenberg
model \cite{Cross/Hohenberg:1993,SwiftHohenberg:1977} symmetric under
the Euclidean group E(2) that allows to study the transition from
higher E(2)xU(1) to lower E(2) symmetry by changing a parameter that
controls the strength of shift symmetry breaking (SSB). Using weakly
nonlinear analysis we derive amplitude equations for stationary planforms
and find three classes of stationary solutions: stripe patterns without
any pinwheels, pinwheel crystals with pinwheels regularly arranged
on a rhombic lattice, and quasi-periodic patterns containing a large
number of irregularly spaced pinwheels. We calculate the phase diagram
of these solutions depending on the strength of SSB, the effective
strength of nonlocal interactions, and the range of nonlocal interactions.
With increasing strength of SSB, pinwheel free patterns are progressively
replaced by pinwheel crystals in the phase diagram while both pinwheel
free patterns and pinwheel crystals remain stable. Phases of aperiodic
pinwheel rich patterns remain basically unaffected. A critical strength
of SSB exists above which multistable aperiodic patterns collapse
into a single aperiodic state. %
\begin{figure}
\includegraphics[clip,width=0.95\columnwidth]{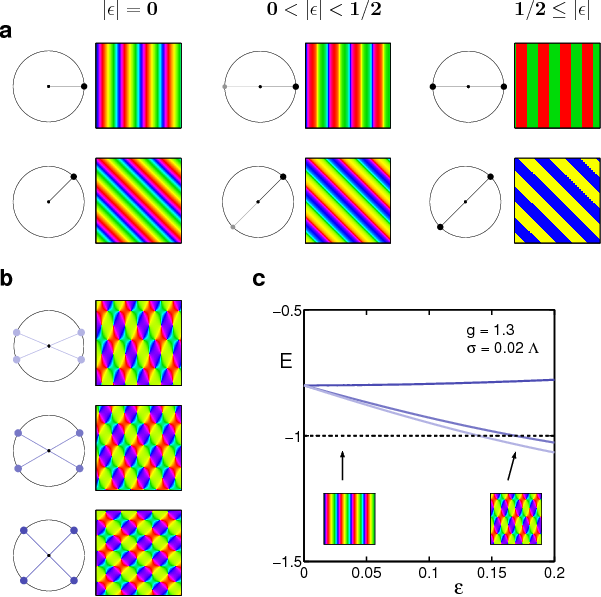}

\caption{\textbf{(a)} Plane waves with wavevector in horizontal (\emph{top})
and oblique (\emph{bottom}) direction for variable strength of SSB
\textbf{$(\epsilon=0,\,0.35,\,1)$ (b)} PWCs of varying intersection
angle $\alpha$, $\pi/4\le\alpha\le\pi/2$. \textbf{(c)} Energy of
solutions depends on $\epsilon$ and on $\alpha$. For sufficiently
large $\epsilon$ PWCs are energetically favored relative to plane
waves. (\emph{dashed}: energy of plane waves, \emph{plain}: energies
of PWCs for $\alpha=\pi/4,\,\pi/3,\,\pi/2$) \label{fig:1}}
\end{figure}

The spatial structure of an OPM can be represented by a complex field
$z(\mathbf{x})$ where $\mathbf{x}$ denotes the 2D position of neurons
in the visual cortex, $\theta(\mathbf{x})=\arg\left(z(\mathbf{x})\right)/2$
their preferred stimulus orientation, and the modulus $|z(\mathbf{x})|$
is a measure of their selectivity \cite{Swindale:1982b}. In this
representation, pinwheel centers are zeros of the field $z(\mathbf{x})$.
The simplest models for the formation of OPMs are defined by a dynamics
\begin{equation}
\partial_{t}z(\mathbf{x},t)=F[z](\mathbf{x},t).\label{eq:1}\end{equation}
where $t$ denotes time and $F[z]$ is a nonlinear operator. We assume
the dynamics equivariant under translation $T_{\mathbf{y}}z(\mathbf{x})=z(\mathbf{x}+\mathbf{y})$,
rotation $R_{\alpha}z(\mathbf{x})=e^{2i\alpha}z(\Omega_{-\alpha}\mathbf{x})$
with rotation matrix $\Omega_{\phi}$, and reflection at the cortical
$(1,0)$ axis $Pz(\mathbf{x})=\bar{z}(\bar{\mathbf{x}})$, thus expressing
the fact that within cortical layers there are no special locations
or directions \cite{Braitenberg/Schuz:1998}. In addition, if interactions
between OPM development and visuotopy are neglected it is also equivariant
under global shifts of orientation preference $S_{\beta}z(\mathbf{x})=e^{i\beta}z(\mathbf{x})$
(\emph{shift symmetry}). Rotations $R$ thus consist of a composition
of phase shifts $S$ and coordinate rotations $D$, i.e. $R_{\alpha}=S_{2\alpha}\circ D_{\alpha}$
with $D_{\alpha}z(\mathbf{x})=z(\Omega_{-\alpha}\mathbf{x})$. 

We consider the general class of variational models \cite{Lee:2003,Wolf:2005b,Swindale:1982b}
for which $F[z]$ has the form 

\begin{equation}
F[z]=Lz+\epsilon M\bar{z}+N_{3}[z].\label{eq:F[z]}\end{equation}
Here $L$ is a linear, translation invariant and self-adjoint operator,
that accounts for a finite wavelength instability. $N_{3}$ is a cubic
nonlinearity which stabilizes the dynamics. The second term involves
a complex conjugation $Cz=\bar{z}$ and thus manifestly breaks shift
symmetry when $\epsilon\neq0$. $M$ is assumed to be linear, translation
invariant and bounded. Equivariance under rotations, $[MC,\, R_{\alpha}]=0$,
requires \begin{equation}
D_{\alpha}MD_{\alpha}^{-1}=S_{-4\alpha}M\label{eq:shtw_trof_rule}\end{equation}
and equivariance under parity $[M,\, P]=0$. 

As a concrete example we will consider the model \begin{eqnarray}
L & = & r-(k_{c}^{2}+\nabla^{2})^{2}\label{eq:L}\\
N_{3}[z] & = & (1-g)|z(\mathbf{x})|^{2}z(\mathbf{x})-\frac{2-g}{2\pi\sigma^{2}}\label{eq:N[z]}\\
 & \times & \int d^{2}y\left(|z(\mathbf{y})|^{2}z(\mathbf{x})+\frac{1}{2}z(\mathbf{y})^{2}\bar{z}(\mathbf{x})\right)e^{-|\mathbf{y}-\mathbf{x}|^{2}/2\sigma^{2}}\nonumber \\
M & = & r{(\partial}_{x}+i\partial_{y})^{4}{(\partial}_{xx}+\partial_{yy})^{-2}\label{eq:M}\end{eqnarray}
where $L$ is the Swift-Hohenberg operator \cite{Cross/Hohenberg:1993,SwiftHohenberg:1977}
with critical wavenumber $k_{c}$ and instability parameter $r$.
$N_{3}$ is adopted from \cite{Wolf:2005b}, where $\sigma$ sets
the range of the nonlocal interactions and $g$ determines whether
the local $(g>1)$ or the nonlocal term $(g<1)$ stabilizes the dynamics.
$M$ is the simplest differential operator which transforms according
to Eq.(\ref{eq:shtw_trof_rule}). It is unitary with spectrum $\propto e^{4i\arg(\mathbf{k})}$. 

We used weakly nonlinear analysis \cite{Manneville:1990} to study
potential solutions of Eq.(\ref{eq:F[z]}). We consider planforms
$z(\mathbf{x})=\sum_{j=0}^{2n-1}A_{j}e^{i\mathbf{k}_{j}\mathbf{x}},\,\,\mathbf{k}_{j}=k_{c}(\cos\alpha_{j},\,\sin\alpha_{j})$
with $2n$ modes where we require that to each mode also its antiparallel
mode is in the set. By symmetry, the dynamics of the amplitudes $A_{j}$
at threshold has the form\begin{equation}
\dot{A}_{j}=A_{j}+\epsilon\bar{A}_{j-}e^{4i\alpha_{j}}-\sum_{k=0}^{2n-1}g_{jk}|A_{k}|^{2}A_{j}-\sum_{k=0}^{2n-1}f_{jk}A_{k}A_{k-}\bar{A}_{j-}\label{eq:amplitude}\end{equation}
where $j^{-}$ denotes the index of the mode antiparallel to mode
$j$, $\mathbf{k}_{j^{-}}=-\mathbf{k}_{j}$ and with real valued and
symmetric matrices $g_{jk}$ and $f_{jk}$ which determine the coupling
and competition between modes. They can be expressed in terms of angle-dependent
interaction functions $g(\alpha)$ and $f(\alpha)$, which are obtained
from the nonlinearity $N_{3}[z]$ (cf.\cite{Wolf:2005b,Cross/Hohenberg:1993,Manneville:1990}).
For simplicity we restrict the following analysis to the class of
permutation symmetric models, defined in \cite{Wolf:2005b}, for which
$g(\alpha)=g(\alpha+\pi)$. %
\footnote{For the nonlinearity Eq.(\ref{eq:N[z]}) one obtains $g(\alpha)=g+(2-g)e(\alpha)$
and $f(\alpha)=\frac{1}{2}g(\alpha)$ with $e(\alpha)=2\exp(-\sigma^{2}k_{c}^{2})\cosh(\sigma^{2}k_{c}^{2}\cos\alpha)$
(cf.\cite{Wolf:2005b} for details). The coupling coefficients are
given by $g_{jk}=(1-\frac{1}{2}\delta_{jk})g(|\alpha_{k}-\alpha_{j}|)$
and $f_{jk}=(1-\delta_{jk}-\delta_{jk^{-}})f(|\alpha_{k}-\alpha_{j}|)$. %
}

The simplest solution to Eq.(\ref{eq:amplitude}) is obtained for
$n=1$ and consists of plane waves with wavevector $\mathbf{k}=k_{c}(\cos\alpha,\sin\alpha)$.
For $|\epsilon|\le1/2$ it is given by \begin{equation}
z(\mathbf{x})=\frac{e^{2i\alpha}}{\sqrt{g_{00}}}[\sqrt{1+2\epsilon}\,\cos(\mathbf{k}\mathbf{x}+\phi)+i\sqrt{1-2\epsilon}\sin(\mathbf{k}\mathbf{x}+\phi)]\label{eq:plane wave}\end{equation}
with arbitrary phase $\phi$. Hence with SSB orientation angles are
no longer equally represented. For $\epsilon>0$ cortical area for
orientations $\alpha$ and $\alpha+\pi/2$ is recruited at the expense
of $\alpha+\pi/4$ and $\alpha+3\pi/4$ (and vice versa for $\epsilon<0$).
Beyond a critical strength of SSB, $\epsilon_{*}=1/2$, patterns only
contain two orientations, $z(\mathbf{x})=e^{2i\alpha}\mathcal{N}\,\cos(\mathbf{k}\mathbf{x}+\phi)$
for $\epsilon>\epsilon_{*}$ and $z(\mathbf{x})=ie^{2i\alpha}\mathcal{N}\,\sin(\mathbf{k}\mathbf{x}+\phi)$
for $\epsilon<-\epsilon_{*}$ with $\mathcal{N=}\sqrt{4(1+|\epsilon|)/3g_{00}}$
(Fig.\ref{fig:1}a). 

Another class of solutions, rhombic pinwheel crystals (PWCs), exist
for $n=2$ and consist of two pairs of antiparallel modes forming
an angle $0<\alpha\le\pi/2$ which are characterized by $|A_{0}|=|A_{0^{-}}|=a=|A_{1}|=|A_{1^{-}}|$.
We consider w.l.o.g. the case $\alpha_{0}=-\alpha/2$ and $\alpha_{1}=\alpha/2$
(Fig.\ref{fig:1}b). With $A_{0,1}=ae^{i\mu_{0,1}},\, A_{0^{-},1^{-}}=ae^{i\nu_{0,1}}$
and $\Sigma_{0,1}:=\mu_{0,1}+\nu_{0,1}$ the stationary state is given
by $\Sigma_{1}=-\Sigma_{0}$ and $a^{2}=(1+\epsilon\cos(\Sigma_{0}+2\alpha))/\zeta$
where $\zeta=3g_{00}+2g_{01}+2f_{01}\cos2\Sigma_{0}$. The phase $\Sigma_{0}$
is the solution to $0=\sin2\Sigma_{0}+\epsilon[\sin(\Sigma_{0}-2\alpha)-(2+3g_{00}/g_{01})\sin(\Sigma_{0}+2\alpha)]$
which bifurcates from $\Sigma_{0}=\pm\pi/2$ for $\epsilon=0$. The
energy is given by $E_{PWC}=-4a�(1+|\epsilon|\cos(\Sigma_{0}+2\alpha))+2a^{4}\zeta$.
For the model Eqs.(\ref{eq:L}-\ref{eq:M}) the $\epsilon$ and $\alpha$
dependence of the energy is shown in Fig.\ref{fig:1}c. The solution
then reads $z(\mathbf{x})=2a\left[e^{i\Sigma_{0}/2}\cos(\mathbf{k}_{0}\mathbf{x}+\Delta_{0}/2)+e^{-i\Sigma_{0}/2}\sin(\mathbf{k}_{1}\mathbf{x}+\Delta_{1}/2)\right]$
with arbitrary $\Delta_{0}$ and $\Delta_{1}$.

\begin{figure}
\includegraphics[clip,width=0.95\columnwidth]{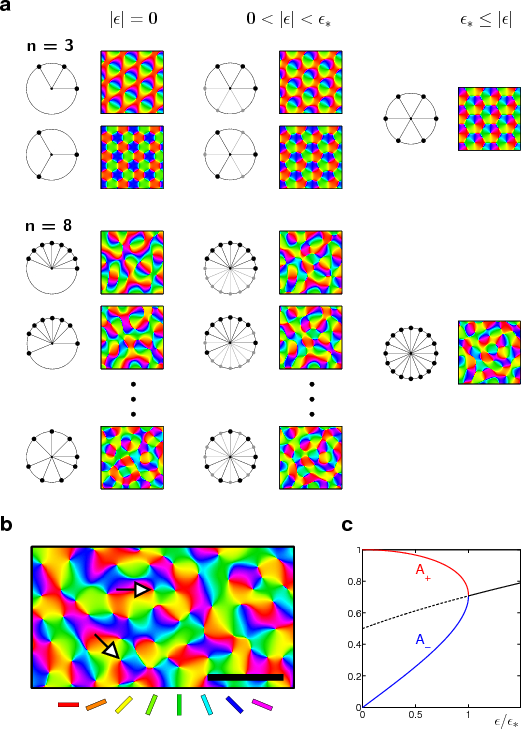}

\caption{Attractors of amplitude equations (Eq.\ref{eq:amplitude}) with full,
partially broken and completely broken shift symmetry. \textbf{(a)}
ECPs. Preferred orientations are color coded {[}see bars in (b)].
Arrangement of active modes on the critical circle and corresponding
OPMs. For n = 3 and 8 there are 2 and 15 different classes of ECPs,
respectively. Complete (partial, no) suppression of opposite modes
for full (weakly broken, maximally broken) shift symmetry \emph{(left,
middle, right column).} \textbf{(b)} OPM in tree shrew V1 (data: L.E.White,
Duke Univ., USA). Arrows pinwheel centers. Scale bar 1 mm. \textbf{(c)}
With increasing degree of symmetry breaking $\epsilon$ amplitudes
of antiparallel modes $A_{-}$ grow and eventually (at $\epsilon=\epsilon_{*}$)
reach the same absolute value as active modes $A_{+}.$\label{fig:2}}
\end{figure}

A large set of quasiperiodic solutions originates from the essentially
complex planforms (ECP) $z(\mathbf{x})=\sum_{j=0}^{n-1}A_{j}^{+}e^{il_{j}\mathbf{k}_{j}\mathbf{x}}$
that solve Eq. (\ref{eq:amplitude}) for $\epsilon=0$ \cite{Wolf:2005b}.
Here, wave vectors $\mathbf{k}_{j}=k_{c}(\cos\frac{\pi}{N}j,\,\sin\frac{\pi}{N}j)$
$(j=0,\dots,\, n-1)$ are distributed equidistantly on the upper half
of the critical circle and binary variables $l_{j}=\pm1$ determine
whether the mode with wave vector $\mathbf{k}_{j}$ or with wave vector
$-\mathbf{k}_{j}$ is active (Fig.\ref{fig:2}(a) \emph{left column}).
\begin{figure}
\includegraphics[clip,width=0.95\columnwidth]{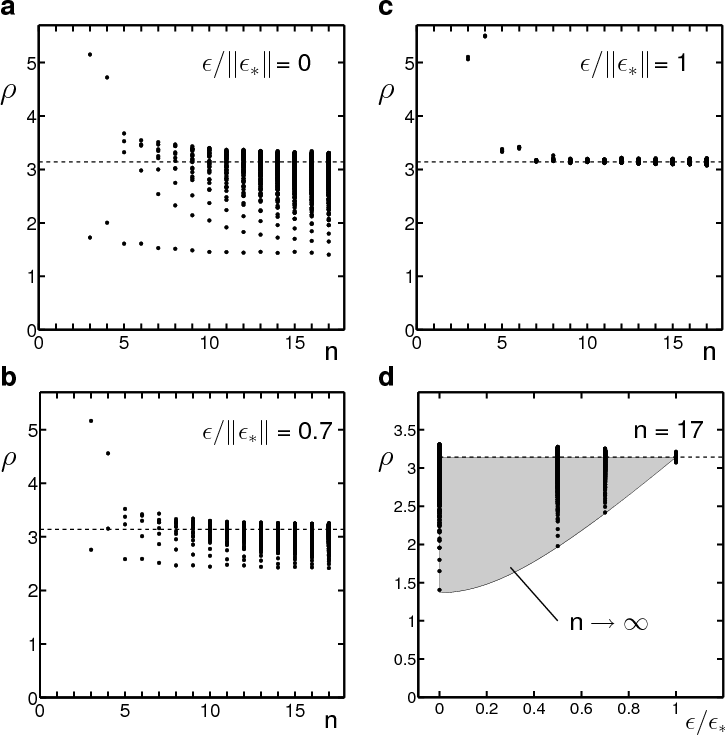}

\caption{\textbf{(a)}-\textbf{(c)} Pinwheel densities for all realizations
of ECPs with $3\le n\le17$ and different degrees of shift symmetry
breaking $\epsilon$. \textbf{(d)} Pinwheel densities for $n=17$
\emph{(dots)} and for $n\to\infty$ in the Gaussian approximation
\emph{(gray region)}.}
\end{figure}

For $|\epsilon|>0$ we find that ECPs generalize to $z(\mathbf{x})=\sum_{j=0}^{n-1}\left[A_{j}^{+}e^{il_{j}\mathbf{k}_{j}\mathbf{x}}+A_{j}^{-}e^{-il_{j}\mathbf{k}_{j}\mathbf{x}}\right]$.
For each $n$ there exists a critical value $\epsilon_{*}:=\gamma{[g}_{00}+2\sum_{j=0}^{n-1}f_{ij}\cos\frac{4\pi}{n}j]$,
where $\gamma=[2\sum_{j=0}^{n-1}g_{ij}]^{-1}$. When $|\epsilon|\le{|\epsilon}_{*}|$
stationary amplitudes $a^{\pm}=|A_{j}^{\pm}|$ fulfill $a_{\pm}^{2}=\gamma(1\pm\sqrt{1-\epsilon^{2}/\epsilon_{*}^{2}})$.
When $|\epsilon|\ge{|\epsilon}_{*}|$ the amplitude of antiparallel
and active modes are equal $a_{\pm}^{2}=\gamma[1+|\epsilon|]/[1+\epsilon_{*}]$
(Fig.\ref{fig:2}c). A simple measure of the degree to which SSB affect
$n$-ECPs is \[
q:=\frac{\sum_{j}A_{j}^{+}A_{j}^{-}e^{-i\frac{4\pi}{n}j}+c.c.}{\sum_{j}\left|A_{j}^{+}\right|^{2}+\left|A_{j}^{-}\right|^{2}}\,.\]
For a stationary $n$-ECP we find $q=\epsilon/|\epsilon_{*}|$ if
$|\epsilon|\le{|\epsilon}_{*}|$ and $q=\mbox{sign}\,\epsilon$ if
$|\epsilon|>{|\epsilon}_{*}|$. 

Stationary phases $\phi_{j}^{\pm}=\arg A_{j}^{\pm}$ fulfill the condition
$\phi_{j}^{+}+\phi_{j}^{-}=(4\pi/n)\, j$ if $\epsilon>0$ and $\phi_{j}^{+}+\phi_{j}^{-}=(4\pi/n)\, j+\pi$
if $\epsilon<0$. This implies that all orientations are represented
in patterns with $n\ge3$. The solution can be written \[
z(\mathbf{x})=\sqrt{2\gamma}\sum_{j=0}^{n-1}[\sqrt{1+q}\, z_{j}^{e}(\mathbf{x},\phi_{j})+\sqrt{1-q}\, z_{j}^{o}(\mathbf{x},\phi_{j})]\]
with $z_{j}^{e}(\mathbf{x},\phi_{j})=e^{i\frac{2\pi}{n}j}\cos(l_{j}\mathbf{k}_{j}\mathbf{x}+\phi_{j})$
and $z_{j}^{o}(\mathbf{x},\phi_{j})=ie^{i\frac{2\pi}{n}j}\sin(l_{j}\mathbf{k}_{j}\mathbf{x}+\phi_{j})$
and arbitrary phases $\phi_{j}$. For $n=1$ this is in agreement
with Eq.(\ref{eq:plane wave}) where $\epsilon_{*}=1/2$. Reflection
at the axis parallel to $\mathbf{k}_{j}$ acts on the functions $z_{j}^{e}$
and $z_{j}^{o}$ as $+1$ and $-1$, respectively. Thus $z_{j}^{e}$
and $z_{j}^{o}$ correspond to the even and odd eigenvectors of the
nullspace of $L+\epsilon MC$ (cf.\cite{Thomas:2004}). For $\epsilon>0$
$(\epsilon<0)$ the even (odd) part dominates the solution. 

The dynamics Eq.(\ref{eq:amplitude}) exhibits a potentially exceedingly
high number of multistable solutions. The energy of $n$-ECPs is given
by $E_{n}=-n\gamma[1+\epsilon^{2}/\epsilon_{*}]$ for $|\epsilon|\le\epsilon_{*}$
and $E_{n}=-n\gamma(1+|\epsilon|)^{2}/[1+\epsilon_{*}]$ for $|\epsilon|\ge\epsilon_{*},$
respectively, and does not depend on the variables $l_{j}$ which
identify a particular $n$-ECP. Due to the growth of antiparallel
modes with increasing $|\epsilon|$ patterns for all different realizations
$l_{j}$ with phases $\phi_{j}:=l_{j}\Phi_{j}+\frac{1}{4}(1-\mbox{sign}(\epsilon))(1-l_{j})\pi$
($\Phi_{j}$ arbitrary but fixed) eventually collapse in a single
state $z(\mathbf{x})\propto\sum_{j=0}^{n-1}z_{j}^{e/o}(\mathbf{x},\phi_{j})$
(Fig.\ref{fig:2}a).%
\begin{figure}
\includegraphics[clip,width=0.95\columnwidth]{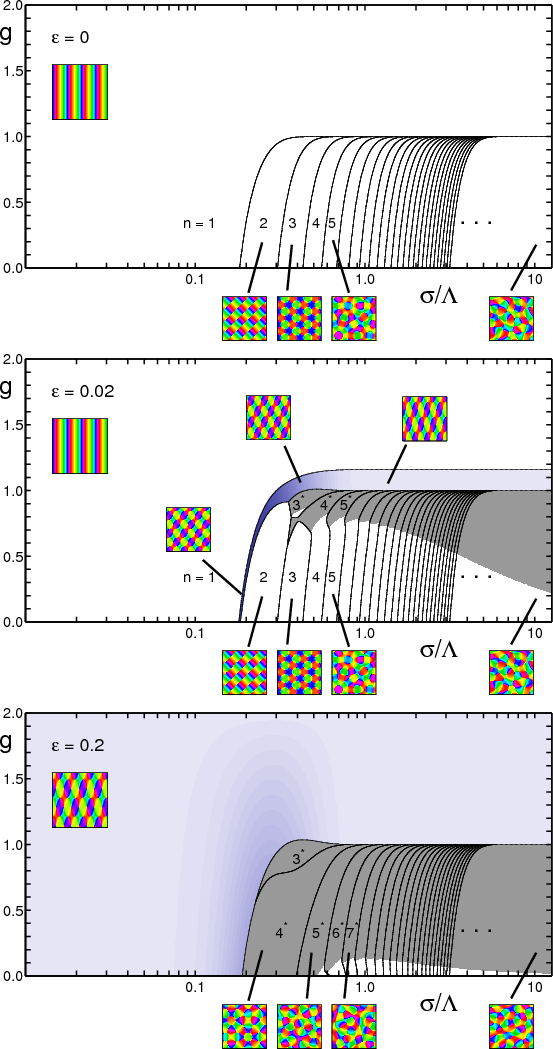}

\caption{Phase diagrams of the model, Eqs.(\ref{eq:L}-\ref{eq:M}), near
criticality for variable SSB $\epsilon$. The graph shows the regions
of the $g-\sigma/\Lambda$ plane in which \emph{n}-ECPs and PWCs have
minimal energy ($n=1-25,\, n>25$ dots). Regions of maximally broken
shift symmetry {[}$\epsilon\ge\epsilon_{*}(N,g,\sigma)$] shaded in
\emph{gray}. Regions where $\alpha$-PWCs prevail is shaded in \emph{blue,}
intensity level codes for the relative angle \emph{$\alpha$. (light
blue: $\pi/4\le\alpha\le\pi/2\,:$dark blue)} \label{fig:Phase-diagrams}}
\end{figure}
\begin{figure}
\includegraphics[clip,width=0.95\columnwidth]{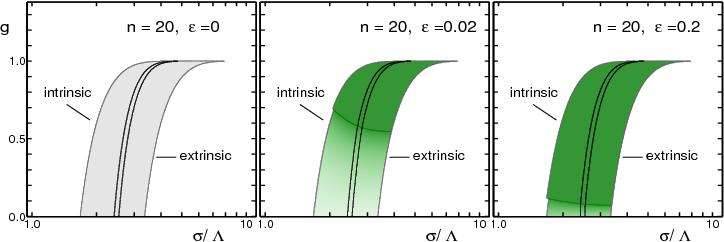}

\caption{Stability regions of ECPs with $n=20$ for different $\epsilon$.
Dashed line denotes the critical line $\epsilon=\epsilon_{*}(N,g,\sigma)$,
above which $q=1$. Shaded region denotes region in the $g-\sigma/\Lambda$
plane for which that planform is a stable solution of the dynamics
and coexists with planforms of nearby values of $n$, e.g. $n=18,\,19,\,21,\,22$.
In the inner region (marked by the two inner lines) this solution
minimizes energy. \label{fig:4}}
\end{figure}

This collapse manifests itself in the pinwheel densities $\rho_{n}$
of various ECPs shown in Fig.3a-c for different strength of SSB. The
pinwheel density of an $n$-ECP in the large $n$ limit is $\rho(\chi)=\pi\sqrt{1-\frac{8}{\pi^{2}}(1-\epsilon^{2}/\epsilon_{*}^{2})\zeta^{2}}$
and depends through $\zeta=\frac{1}{4n}|\sum_{j=0}^{n-1}l_{j}\mathbf{k}_{j}|\le1$
on the configuration of active modes. Fig.3d shows that pinwheel densities
fill a band of values. With increasing degree of SSB this band narrows
and pinwheel densities eventually equal $\pi$ at the critical value
$|\epsilon|=\epsilon_{*}.$ 

To reveal how SSB affects pattern selection we calculated the phase
diagram for the model specified in Eqs.(\ref{eq:L}-\ref{eq:M}) for
various values of $\epsilon$. Fig.\ref{fig:Phase-diagrams} shows
the configurations of $n$-ECPs and PWCs minimizing the energy %
\footnote{The dynamics Eq. (\ref{eq:amplitude}) has the real valued energy
functional $E=\sum_{j}|A_{j}|�+\epsilon\sum_{j}(\bar{A}_{j}\bar{A}_{j-}e^{4i\frac{\pi}{n}j}+c.c.)-\frac{1}{2}\sum_{jk}g_{jk}|A_{j}|^{2}|A_{k}|^{2}-\frac{1}{2}\sum_{jk}f_{jk}A_{k}A_{k-}\bar{A}_{j}\bar{A}_{j-}$.%
}. Plane waves are progressively replaced by $\alpha$-PWCs with increasing
SSB. Depending on the location in parameter space and on $\epsilon$,
a particular angle $\alpha$ minimizing the energy (c.f. Fig.\ref{fig:1}c).
Large $n-$ECPs are selected when the dynamics is stabilized by long-range
interactions ($g<1$, $\sigma>\Lambda$). In this parameter regime
plane waves and pinwheel crystals are unstable. The degree of SSB
$q$ manifest in a given $n-$ECP attractor depends on $\epsilon$
and on the location in the phase diagram. Above a critical line defined
by $|\epsilon_{*}(N,g,\sigma)|=|\epsilon|$ antiparallel modes are
maximal and $|q|=1$ (\emph{gray area}), below that line $|q|\le1$.
Figs.~\ref{fig:Phase-diagrams} and \ref{fig:4} show the high sensitivity
of the dynamics to even small amounts of SSB, a substantial area in
phase is occupied by ECPs with $|q|=1$ even for $\epsilon=0.02$. 

Our analysis of pattern selection in visual cortical development demonstrates
that dynamical models of orientation map development are very sensitive
to the presence of interactions imposed by Euclidean E(2) symmetry.
It also reveals that the impact of the Euclidean symmetry of perceptual
visual space is not an all or none phenomenon. For weak SSB our Euclidean
model closely mimics the behavior of models possessing the higher
E(2)xU(1) symmetry. The only qualitative change that we found in the
transition from higher to lower symmetry was the collapse of multistable
solutions. This collapse, however, only happens when a finite critical
strength of SSB is reached. Up to this threshold strength, models
exhibiting E(2) and E(2)xU(1) symmetry seem to be topologically conjugate
to one another.

Our analysis also reveals that the impact of Euclidean symmetry differs
qualitatively for aperiodic and periodic patterns. For aperiodic patterns,
(i) the parameter regime in which they possess minimal energy is virtually
unaffected by the strength of SSB, (ii) sets of different multistable
solutions become progressively more similar and finally merge forming
a single unique ground state, and (iii) Euclidean symmetry geometrically
manifests itself in specific two point correlations that are absent
in the E(2)xU(1) symmetric limit. For periodic patterns such as pinwheel
crystals and pinwheel free states, (i) the parameter regime in which
these solutions possess minimal energy depends on the strength of
SSB, and (ii) Euclidean symmetry geometrically manifests itself in
a specifically selected tilt angle of rhombic pinwheel crystals and
a wave vector dependent underrepresentation of particular orientations
for pinwheel free states.

The Swift-Hohenberg model of Euclidean symmetry considered here predicts
that aperiodic pinwheel rich patterns resembling the architecture
of the primary visual cortex are only stable when long-range interactions
dominate pattern selection, confirming previous predictions of a model
of higher E(2)xU(1) symmetry. Our analysis predicts that in this regime,
two point correlations provide a sensitive measure of the strength
of SSB. Such correlations are therefore a promising tool for probing
the impact of Euclidean symmetry in visual cortex development. \bibliographystyle{apsrev}
\bibliography{bib}

\end{document}